\documentclass[a4paper,12pt]{amsart}
\usepackage{amssymb,amsmath,latexsym,color,graphicx}
\usepackage{hyperref}
\usepackage{amsfonts}
\usepackage{amsthm}
\newtheorem{dref}{Definition}[section] 
\newtheorem{theo}[dref]{Theorem} 

\newtheorem{remark}[dref]{Remark} 

\newtheorem{cor}[dref]{Corollary}
\newtheorem{conj}[dref]{Conjecture}

\begin{document}

\title[]{Higher order degenerations of Fay's identities and 
applications to integrable equations}
\author{Christian Klein}
\address{Institut de Math\'ematiques de Bourgogne, UMR 5584\\
                Universit\'e de Bourgogne-Franche-Comt\'e, 9 avenue Alain Savary, 21078 Dijon
                Cedex, France\\
				Institut Universitaire de France\\
    E-mail Christian.Klein@u-bourgogne.fr}
    
\author{Jordi Pillet}
\address{Institut de Math\'ematiques de Bourgogne, UMR 5584\\
                Universit\'e de Bourgogne-Franche-Comt\'e, 9 avenue Alain Savary, 21078 Dijon
                Cedex, France\\
    E-mail jordi.pillet@u-bourgogne.fr}
	\address{Department of Mathematics and Statistics, Concordia University, 1455 de Maisonneuve W., Montreal, QC H3G 1M8, Canada.}
    
\date{\today}
\begin{abstract}
	Higher order degenerated versions of Fay's trisecant identity are 
	presented. It is shown that these lead to solutions for  
	Schwarzian Kadomtsev-Petviashvili equations. 

\end{abstract}

\thanks{This work is partially supported by  the isite BFC, the EIPHI Graduate School (contract ANR-17-EURE-0002) and by the 
European Union Horizon 2020 research and innovation program under the 
Marie Sklodowska-Curie RISE 2017 grant agreement no. 778010 IPaDEGAN. 
We thank M.~Pavlov for helpful discussions and hints.}

\maketitle

\section{Introduction}
Solutions to integrable partial differential equations (PDEs) in 
terms of multi-dimensional theta functions on compact Riemann 
surfaces appeared in the 1970s in the search for quasi-periodic 
solutions, see for instance \cite{Dub,BBEIM} for a historic account. 
These solutions were constructed via the \emph{Baker-Akhiezer} 
function, a function with an essential singularity on the Riemann 
surface first introduced by Clebsch and Gordan. Mumford and coworkers 
introduced in \cite{Mum} a complementary approach based on Fay's 
celebrated \textit{trisecant identity} for theta functions 
\cite{fay},
\begin{equation*}
	\Theta^{*}_{ad}\Theta^{*}_{cb}\Theta_{ac}\Theta_{bd}
	+\Theta^{*}_{ca}\Theta^{*}_{db}\Theta_{bc}\Theta_{ad}
	= \Theta^{*}_{cd}\Theta^{*}_{ab}\Theta\Theta_{a+b,c+d},
\end{equation*}
where we have introduced the notation 
\begin{equation}
	\Theta^{*}_{ab}=\Theta^{*}\left(\int_{a}^{b}\right),\quad 
	\Theta_{ab}=\Theta\left(\mathrm{z}+\int_{a}^{b}\right)
	\label{notation};
\end{equation}
here $\Theta(\mathrm{z})$, $\mathrm{z}\in \mathbb{C}^{g}$, 
is the $g$-dimensional Riemann theta function, $\Theta^{*}(\mathrm{z})$ is 
a theta function with an odd non-singular characteristic, see the 
definitions  (\ref{theta}), (\ref{thchar}), 
$\Theta=\Theta(\mathrm{z})$, $\Theta^{*}=\Theta^{*}(0)=0$, and 
$a,b,c,d$ are points on a 
Riemann surface $\mathcal{R}$ with genus $g$. The Abel map 
$\int_{a}^{b}$ between two points $a$ and $b$ on $\mathcal{R}$ is 
defined at the beginning of section 2. Note that the name trisecant 
identity refers to secants on the so-called Kummer variety, see 
\cite{Tai} for a comprehensive review. 

Since Fay's identity (\ref{Fay}) holds for arbitrary points $a$, $b$, 
$c$, $d$ on the  Riemann surface $\mathcal{R}$, it is possible to 
consider the identity in the limit that two or more points 
coincide\footnote{Note that there are generalizations of Fay's 
identity to more than 4 points and degenerations thereof, see for 
instance \cite{Fay2,Dub2,Ber} and references therein.}. 
This leads to identities between derivatives of theta functions making 
it possible to identify solutions to certain PDEs from degenerated 
identies. In \cite{Mum} this was done for the Sine-Gordon equation 
and the Kadomtsev-Petviasvili (KP) equation. On special Riemann surfaces 
(hyperelliptic, trigonal) the latter solutions lead to 
algebro-geometric solutions for the Korteweg-de Vries (KdV) \cite{Mum} and 
the Boussinesq equation \cite{BBEIM}. In \cite{KKS} previously known solutions to the 
Ernst equation \cite{Kor} were reconstructed via Fay's identity, see 
also \cite{ernstbook}, in \cite{KKch} 
known 
solutions to the Camassa-Holm equation \cite{GH} were obtained with Mumford's 
approach. In \cite{KalIMRN} Kalla 
presented a new degenerated identity allowing to identify known
solutions to the nonlinear Schr\"odinger \cite{Its,Pre} and Davey-Stewartson 
equations \cite{Mal} and to construct solutions to vector nonlinear 
Schr\"odinger equations in terms of theta functions. For a recent 
review on completely integrable dispersive PDEs, we refer to 
\cite{book}.  In this paper we generalize Kalla's approach 
to higher order in the local parameter near the point $a$. We obtain 
with the above notation\\
\textbf{Main theorem Part I}\\
	Let $a$, $b$ be points on a compact Riemann surface $\mathcal{R}$,  
	and let $U:=D_{b}\ln \Theta 
	\Theta^{*}_{ba}$. Then $U$ satisfies
\begin{equation}
	\begin{split}
		0&=2(D_{a}U)^2 D_{a}''D_{a}U-2D_{a}U D_{a}''UD_{a}^{2}U
		-(D_{a}U)^2 D_{a}^{4}U+4D_{a}UD_{a}^{3}U D_{a}^{2}U\\
		&-3(D_{a}^{2}U)^{3}+3(D_{a}'U)^{2}D_{a}^{2}U
		-3(D_{a}U)^2(D_{a}')^{2}DU.
\end{split}
	\label{pillet3}
\end{equation}

This identity has similarities to the classical identity (\ref{Fay3}) 
by Fay in the sense that it involves the derivatives $D_{a}''$, 
$D_{a}'$ and $D_{a}$ of $\Theta(\mathrm{z})$, but appears to be new. 
In contrast to the potential in (\ref{Fay3}), the function $U$ also 
depends on a point $b$ on the Riemann surface $\mathcal{R}$ which is 
distinct from $a$, but otherwise 
arbitrary.

We also prove \\
\textbf{Main theorem Part II}\\
The function \( \phi(x,y,t) := \mathrm{D}_b \ln{\Theta^*_{ab} \Theta 
(x\mathbf{v}_{0}(a)+y \mathbf{v}_{1}(a)+ t 
\mathbf{v}_{2}(a)+\mathbf{d} }) \), $(x,y,t)\in \mathbb{R}^{3}$,  solves the Schwarzian KP equation:
\begin{equation}
    \Big( \frac{\phi_t}{\phi_x}-\frac{1}{2} \{ \phi ; x \} \Big)_x - \frac{3}{2} \Big( \frac{\phi_y}{\phi_x} \Big)_y-\frac{3}{4} \Big( \frac{\phi_y^2}{\phi_x^2} \Big)_x=0
    \label{SKP}
\end{equation}
where \( \{ \phi ; x \} \) denotes the Schwarzian derivative along \( 
x \): \( \{ \phi ; x \} := \frac{\phi_{xxx}}{\phi_x}-\frac{3}{2} 
\Big( \frac{\phi_{xx}}{\phi_x}\Big)^2 \), where  
the indices denote partial derivatives with respect to the respective 
variable, and where $\mathbf{v}_{j}$, $j=0,1,2$ has the components 
$v_{ij}$, $i=1,\ldots,g$ defined in (\ref{abelexp}). 

The solution $\phi$ in terms of multi-dimensional theta functions for 
the Schwarzian KP equation seems to be new. 
The Schwarzian KP equation (\ref{SKP}) appeared first in the 
Painlev\'e analysis of the KP equation in \cite{WeissI} as a singularity manifold equation. Its 
integrability was established in \cite{BK}. As in the case of the KP 
equation, a reduction to a Schwarzian KdV and Boussinesq equation is 
possible.

The paper is organised as follows: in section 2 we collect some basic 
definitions of quantities defined on a compact Riemann surface and 
known facts on Fay's identities. In section 3 we rederive identities 
(\ref{Fay2}) and (\ref{kalla1}) from identity (\ref{Fay1}) and prove 
the first part of the main theorem. In section 4 this is applied to 
integrable PDEs. We add some concluding remarks in section 5. 

\section{Preliminaries}
In this section, we will collect some basic definitions and known 
facts on Fay's identities and applications. 

\subsection{Basic definitions}

In this paper we always consider a Riemann surface $\mathcal{R}$ of 
genus $g\in \mathbb{N}$ equipped with a canonical basis of cycles
$a_{1},\ldots,a_{g}, b_{1},\ldots,b_{g}$
satisfying the intersection conditions
$$    a_{i}\circ b_{j}=\delta_{ij},\quad a_{i}\circ a_{j}=0,\quad b_{i}\circ 
    b_{j}=0, \quad i,j=1,\ldots,g.
$$ The $g$-dimensional vector of holomorphic 
1-forms is denoted by $\mathrm{d}\omega$ and normalized by 
$\int_{a_{i}}^{}\mathrm{d}\omega_{j}=\delta_{ij}$, $i,j=1,\ldots,g$. 
The matrix of $b$-periods 
$\mathbb{B}_{ij}=\int_{b_{i}}^{}\mathrm{d}\omega_{j}$, 
$i,j=1,\ldots,g$, is a Riemann matrix, i.e., it is symmetric and has 
a positive definite imaginary part. 
The Abel map $\omega:P\mapsto \int_{P_{0}}^{P}\mathrm{d}\omega$ 
is a bijective map from the Riemann surface $\mathcal{R}$ into the \emph{Jacobian} 
$Jac(\mathcal{R}):=\mathbb{C}^{g}/\Lambda$ where $\Lambda$ is the 
lattice formed by the periods of the holomorphic 1-forms, 
$$\Lambda=\left\{\mathrm{m}+\mathbb{B}\mathrm{n}: 
m,n\in\mathbb{Z}^{g}\right\}.$$ The expansion of the Abel map at a point 
$P\in\mathcal{R}$ near a point $a\in\mathcal{R}$ is 
written in the form,
\begin{equation}
    \omega_{i}(P)=\sum_{j=0}^{\infty}v_{ij}\frac{\tau^{j}}{j!},\quad i=1,\ldots,g
    \label{abelexp},
\end{equation}
where $\tau$ is a local parameter in the vicinity of $a$ 
containing also $P$. We define the derivatives acting on a function 
$f(z)$, $z\in \mathbb{C}^{g}$ as
\begin{equation}
	\begin{split}
			D_{a}&:= \sum_{i=1}^{g}v_{i0}\partial_{z_{i}},\quad
	D_{a}':= \sum_{i=1}^{g}v_{i1}\partial_{z_{i}},\quad
	D_{a}'':= \sum_{i=1}^{g}\frac{v_{i2}}{2}\partial_{z_{i}},\\
	D_{a}^{(n)}&:= 
	\sum_{i=1}^{g}\frac{v_{in}}{n!}\partial_{z_{i}},\quad n \in 
	\mathbb{N}.
	\end{split}
	\label{Ddef}
\end{equation}

Multi-dimensional theta functions are the building blocks of  meromorphic functions
on Riemann surfaces. The theta
    function with characteristic     $\left 
	[\mathrm{p},\mathrm{q}\right]$ is defined
    as an infinite series,
    \begin{equation}\label{theta}
    \Theta_{\mathrm{p}\mathrm{q}}(\mathrm{z},\mathbb{B})=
    \sum\limits_{\mathrm{N}\in\mathbb{Z}^g}\exp\left\{
    \mathrm{i}\pi\left\langle\mathbb{B}\left(\mathrm{N}+\mathrm{p}\right),
    \mathrm{N}+\mathrm{p}
    \right\rangle+2\pi \mathrm{i}
    \left\langle \mathrm{z}+\mathrm{q},\mathrm{N}+\mathrm{p}
    \right\rangle\right\}
    \;,
    \end{equation}
    with $\mathrm{z}\in\mathbb{C}^g$ and $\mathrm{p}$, $\mathrm{q}\in{
    \mathbb{R}}^g$, where 
    $\left
    \langle\cdot,\cdot\right\rangle$ denotes the Euclidean scalar product
    $\left\langle \mathrm{N},\mathrm{z}\right\rangle=\sum_{i=1}^gN_iz_i$.
The properties of the Riemann matrix ensure that the series converges
absolutely and that the theta function is an entire function on
$\mathbb{C}^{g}$. A characteristic is called \emph{singular} if the
corresponding theta function vanishes identically. 
Half-integer 
characteristics with $2\mathrm{p},2\mathrm{q}\in \mathbb{Z}^{g}$ are 
called \emph{even} if $4\langle 
\mathrm{p},\mathrm{q}\rangle=0\mbox{ mod } 2$ and \emph{odd} 
otherwise.
Theta functions with odd (even) characteristic are odd
(even) functions of the argument $\mathrm{z}$.  The theta function with
characteristic is related to the Riemann theta function $\Theta$, the
theta function with zero characteristic $\Theta:= \Theta_{\mathrm{00}}$,
via
\begin{equation}
    \Theta_{\mathrm{pq}}(\mathrm{z},\mathbb{B})=\Theta(\mathrm{z}
    +\mathbb{B}\mathrm{p} + \mathrm{q})\exp\left\{\mathrm{i}\pi
    \left\langle\mathbb{B}\mathrm{p},\mathrm{p}\right\rangle+
    2\pi \mathrm{i}\left\langle\mathrm{p},\mathrm{z} + \mathrm{q}\right\rangle
    \right\}\;.
    \label{thchar}
\end{equation}
A theta function with a nonsingular half-integer
characteristic is denoted by $\Theta^{*}$.

\subsection{Fay's identities}

Theta functions on Jacobians satisfy  Fay's celebrated trisecant identity
\cite{fay}. It can be seen as a generalization of the classical 
relation between cross ration functions 
for four arbitrary points $a$, $b$, $c$, $d$ in the euclidean 
plane 
\begin{equation}
    \lambda_{abcd}=\frac{\Theta^{*}(\smallint_{a}^{b})\Theta^{*}(\smallint_{c}^{d})}{
    \Theta^{*}(\smallint_{a}^{d})\Theta^{*}(\smallint_{c}^{b})}
    \label{eq:cross3}\;,
\end{equation}
which is a function on $\mathcal{R}$ that vanishes for $a=b$
and $c=d$ and has poles for $a=d$ and $b=c$.
\begin{theo}[Fay \cite{fay}]
Let $a$, $b$, $c$, $d$ be four points on the Riemann 
surface 	$\mathbb{R}$. Then with the above 
definitions
the following identity     holds
\begin{equation}
      \begin{aligned}
        \lambda_{cabd}\,\Theta(\mathrm{z}+\smallint_{b}^{c})\,
        &\Theta(\mathrm{z}+\smallint_{a}^{d}) +\lambda_{cbad}\,
        \Theta(\mathrm{z}+ \smallint_{a}^{c})\,\Theta(\mathrm{z}+
        \smallint_{b}^{d})\\
       & =\Theta(\mathrm{z})\;\Theta(\mathrm{z}+\smallint_{b}^{c}+
        \smallint_{a}^{d})\;,
      \end{aligned}
    \label{Fay}
\end{equation} 
   $\forall\mathrm{z}\in \mathbb{C}^{g}$. The 
   integration paths in (\ref{Fay}) have to be chosen in a way not to 
   intersect the canonical cycles. 
\end{theo}

Degenerated versions  of  Fay's 
identity lead to identities for derivatives of 
theta functions. In the limit $d\to b$, one finds for (\ref{Fay})
\begin{cor}[Fay \cite{fay}]	
	Let $a$, $b$, $c$ be points on the Riemann surface 
	$\mathcal{R}$. Then the following identity holds,
\begin{equation}
    D_{b}\ln \frac{\Theta(\mathrm{z}+
    \smallint_{a}^{c})}{\Theta(\mathrm{z})}=p_{1}(a,b,c)+p_{2}(a,b,c)
    \frac{\Theta(\mathrm{z}+
    \smallint_{b}^{c})\Theta(\mathrm{z}+
    \smallint_{a}^{b})}{\Theta(\mathrm{z}+
    \smallint_{a}^{c})\Theta(\mathrm{z})}
    \label{Fay1},
\end{equation}
where 
\begin{align}
    p_{1}(a,b,c) &=  D_{b}\ln \frac{\Theta^{*}(
    \smallint_{a}^{b})}{\Theta^{*}(\smallint_{c}^{b})},
    \nonumber\\
    p_{2}(a,b,c) & =\frac{\Theta^{*}(
    \smallint_{a}^{c})D_{b}\Theta^{*}}{\Theta^{*}(
    \smallint_{b}^{c})\Theta^{*}(
    \smallint_{b}^{a})}
    \label{pi}.
\end{align}
\end{cor}

In the limit $c\to a$, equation (\ref{Fay1}) yields
\begin{cor}[Fay \cite{fay}]	
		Let $a$, $b$ be points on the Riemann surface 
	$\mathcal{R}$. Then the following identity holds,
\begin{equation}
    D_{a}D_{b}\ln \Theta(\mathrm{z})=q_{1}(a,b)+q_{2}(a,b)
    \frac{\Theta(\mathrm{z}+
    \smallint_{b}^{a})\Theta(\mathrm{z}+
    \smallint_{a}^{b})}{\Theta(\mathrm{z})^{2}}
    \label{Fay2},
\end{equation}
where 
\begin{align}
    q_{1}(a,b) &=  D_{a}D_{b}\ln \Theta^{*}(
    \smallint_{a}^{b}),
    \nonumber\\
    q_{2}(a,b) & =\frac{D_{a}\Theta^{*}D_{b}\Theta^{*}}{\Theta^{*}(
    \smallint_{a}^{b})^{2}}
    \label{qi}.
\end{align}
\end{cor}

In the limit $b\to a$, equation (\ref{Fay2}) can be cast into 
the form (we suppress the index $a$ at the derivatives)
\begin{cor}[Fay \cite{fay}]	
	The following identity holds on the Riemann surface $\mathcal{R}$,
\begin{equation}
    D^{4}\ln\Theta(\mathrm{z})+6(D^{2}\ln 
    \Theta(\mathrm{z}))^{2}+3D'D'\ln \Theta(\mathrm{z})-2DD''\ln 
    \Theta(\mathrm{z})+c_{1}D^{2}\ln \Theta(\mathrm{z}) +c_{2}=0
    \label{Fay3},
\end{equation}
where 
\begin{equation}
    c_{1}=2\frac{D''\Theta^{*}}{\Theta^{*}}-4\frac{D^{3}\Theta^{*}}{D\Theta^{*}}
    -3\left(\frac{D'\Theta^{*}}{D\Theta^{*}}\right)^{2};
    \label{c1}
\end{equation}
the constant $c_{2}$ 
can be obtained by expanding $q_{1}$ and $q_{2}$ 
(\ref{qi}) in the considered limit to fourth order in the local 
parameter near $a$. 
\end{cor}

Kalla generalized the identity (\ref{Fay2}) to
\begin{theo}[Kalla \cite{KalIMRN}]\label{Ktheorem}
	Let $a$, $b$ be points on $\mathcal{R}$, then the following 
	identity holds,
	\begin{equation}
		\begin{split}
0=&D_{a}'\ln 
\frac{\Theta(\mathrm{z}+\smallint_{a}^{b})}{\Theta(\mathrm{z})} 
+D_{a}^{2}\ln 
\frac{\Theta(\mathrm{z}+\smallint_{a}^{b})}{\Theta(\mathrm{z})}
+\left(D_{a}\ln 
\frac{\Theta(\mathrm{z}+\smallint_{a}^{b})}{\Theta(\mathrm{z})}-K_{1}(a,b)\right)^{2}\\
&+2D_{a}^{2}\ln \Theta(\mathrm{z})+K_{2}(a,b)			
		\end{split}
		\label{kalla1}
	\end{equation}
	where $K_{1}(a,b)$, $K_{2}(a,b)$ depend on the points $a$, $b$, 
	but not on $\mathrm{z}$. 
\end{theo}

\section{Proof of Main theorem Part I}
In this section we will consider various identities following from 
Fay's identity (\ref{Fay1}) in the limit $c\to a$. First we 
identify the known relations (\ref{Fay2}) and (\ref{kalla1}), then we 
prove the first part of the main theorem. 

To this end we write (\ref{Fay1}) in the form 
\begin{equation}
	V^{2}D_{b}
	\frac{V(c)}{V}=D_{b}\Theta^{*}\Theta^{*}_{ac}\Theta_{ab}\Theta_{bc},
	\label{Fay1V}
\end{equation}
where we have put 
\begin{equation}
	V(c) = \Theta_{ac}\Theta^{*}_{cb},\quad V = V(a)=\Theta\Theta^{*}_{ab}.
	\label{V}
\end{equation}

\subsection{Known identities}

For identity (\ref{Fay1V})) we consider a Taylor expansion in the limit $c\to a$ in the local 
parameter $\tau$ in (\ref{abelexp}). In lowest 
order we get (\ref{Fay2}) in the form
\begin{equation}
	D_{a}D_{b}\ln V = 
	\frac{1}{V^{2}}D_{b}\Theta^{*}D_{a}\Theta^{*}\Theta_{ab}\Theta_{ba}
	\label{Fay2V}.
\end{equation}

In  order $\tau^{2}$, we get for (\ref{Fay1V}) 
\begin{equation*}
	\frac{V^{2}}{2}D_{b}(D_{a}^{2}\ln V+D_{a}'\ln V+(D_{a}\ln V)^{2})=
	D_{b}\Theta^{*}D_{a}\Theta^{*}\Theta_{ab}\Theta_{ba}\left(\frac{D_{a}'\Theta^{*}}{2D_{a}\Theta^{*}}
	+D_{a}\ln \Theta_{ba}\right),
\end{equation*}
which can be written with (\ref{Fay2V}) in the form
\begin{equation}
	\frac{1}{2}D_{b}(D_{a}^{2}\ln V+D_{a}'\ln V+(D_{a}\ln V)^{2})=
	D_{a}D_{b}\ln V\left(\frac{D_{a}'\Theta^{*}}{2D_{a}\Theta^{*}}
	+D_{a}\ln \Theta_{ba}\right)
	\label{kalla2V}.
\end{equation}
Since this identity holds for all $\mathrm{z}\in\mathbb{C}$, it also holds 
for $\mathrm{z}$ replaced by $\mathrm{z}+\int_{a}^{b}$. This means $\Theta\mapsto 
\Theta_{ab}$, $\Theta_{ba}\mapsto \Theta$. As shown by Kalla 
\cite{KalIMRN}, the difference between 
identity (\ref{kalla2V}) and (\ref{kalla2V}) after this shift of $\mathrm{z}$ 
reads
\begin{equation}
	\begin{split}
		0=&\frac{1}{2}D_{b}\left(D_{a}'\ln \frac{\Theta_{ab}}{\Theta}
		+\frac{D_{a}^{2}\Theta_{ab}}{\Theta_{ab}}-\frac{D_{a}^{2}\Theta}{\Theta}
		+2D_{a}\ln 
		\Theta^{*}_{ba}D_{a}\ln \frac{\Theta_{ab}}{\Theta}\right)\\
		&+D_{a}D_{b}\ln \Theta\Theta^{*}_{ab}D_{a}\ln 
		\Theta_{ba}
		-D_{a}D_{b}\ln \Theta_{ab}\Theta^{*}_{ab}D_{a}\ln 
		\Theta -\frac{D_{a}'\Theta^{*}}{2D_{a}\Theta^{*}}D_{a}D_{b}\ln \frac{\Theta_{ab}}{\Theta}
	\end{split}
	\label{kalla3}.
\end{equation}
With (\ref{Fay2}), we get
\begin{equation}
	\begin{split}
		0=&\frac{1}{2}D_{b}\left(D_{a}'\ln \frac{\Theta_{ab}}{\Theta}
		+D_{a}^{2}\ln\frac{\Theta_{ab}}{\Theta}+
		\left(D_{a}\ln\frac{\Theta_{ab}}{\Theta}\right)^{2}
		+2D_{a}^{2}\ln \Theta \Theta^{*}_{ba}\right)
		\\
		&+\left(D_{a}\ln 
		\Theta^{*}_{ba}-\frac{D_{a}'\Theta^{*}}{2D_{a}\Theta^{*}}\right)
		D_{a}D_{b}\ln \frac{\Theta_{ab}}{\Theta}
	\end{split}
	\label{kalla4}.
\end{equation}
Introducing the derivative $\nabla_{b}:= 
\sum_{i=1}^{g}v_{i0}(b)\partial_{\mathrm{z}_{i}}$ acting only on $\mathrm{z}$, we 
can write (\ref{kalla4}) in the form
\begin{equation}
	\begin{split}
		0=&\nabla_{b}\left\{\frac{1}{2}\left(D_{a}'\ln \frac{\Theta_{ab}}{\Theta}
		+D_{a}^{2}\ln\frac{\Theta_{ab}}{\Theta}\right)
		+\left(D_{a}\ln 
		\Theta^{*}_{ba}-\frac{D_{a}'\Theta^{*}}{2D_{a}\Theta^{*}}\right)
		D_{a}\ln \frac{\Theta_{ab}}{\Theta}\right.\\
		&\left.+D_{a}^{2}\ln \Theta\Theta^{*}_{ab}
		+\frac{1}{2}\left(D_{a}\ln 
		\frac{\Theta_{ab}}{\Theta}\right)^{2}
\right\}
\end{split}
	\label{kalla5}.
\end{equation}
This implies that relation (\ref{kalla1}) holds with 
\begin{equation}
	\begin{split}
		K(a,b)=&D_{a}'\ln \frac{\Theta_{ab}}{\Theta}
		+D_{a}^{2}\ln\frac{\Theta_{ab}}{\Theta}
		+2\left(D_{a}\ln 
		\Theta^{*}_{ba}-\frac{D_{a}'\Theta^{*}}{2D_{a}\Theta^{*}}\right)
		D_{a}\ln \frac{\Theta_{ab}}{\Theta}\\
		&+2D_{a}^{2}\ln \Theta\Theta^{*}_{ab}
		+\left(D_{a}\ln 
		\frac{\Theta_{ab}}{\Theta}\right)^{2}
\end{split}
	\label{kalla6},
\end{equation}
where $K(a,b)$ just depends on $a$, $b$, but not on $\mathrm{z}$. It 
can be computed for instance by putting $\mathrm{z}=0$ on the right 
hand side of (\ref{kalla6}). This reproduces the proof of Theorem 
\ref{Ktheorem} from \cite{KalIMRN}. 

\subsection{Third order in $\tau$}
In third order of the local parameter $\tau$ we get for (\ref{Fay1})
\begin{equation}
	\begin{split}
		0=&D_{b}\left(\frac{1}{6}D_{a}''\ln V+\frac{1}{2}D_{a}D_{a}' 
		\ln V + \frac{1}{2}D_{a}'\ln V D_{a}\ln 
		V+\frac{1}{6}D_{a}^{3}\ln V \right.\\
		&\left.+\frac{1}{2}D_{a}^{2}\ln V 
		D_{a}\ln V+\frac{1}{6}(D_{a}\ln V)^{3}\right)\\
		&-D_{a}D_{b}\ln V
		\left(\frac{D_{a}''\Theta^{*}}{6D_{a}\Theta^{*}}
		+\frac{D_{a}^{3}\Theta^{*}}{6D_{a}\Theta^{*}}
		+\frac{D_{a}'\Theta^{*}}{2D_{a}\Theta^{*}}
		D_{a}\ln \Theta_{ba}+\frac{1}{2}D_{a}'\ln 
		\Theta_{ba}+\frac{D_{a}^{2}\Theta_{ba}}{2\Theta_{ba}}\right),
	\end{split}
	\label{pillet1aV}
\end{equation}
where we have used (\ref{Fay2V}). With (\ref{kalla2V}), we can 
replace $\Theta_{ba}$ in (\ref{pillet1aV}) to obtain a relation
only involving $V$. To this end we put
\begin{equation}
	F:= \frac{1}{2}D_{b}\left(D_{a}'\ln 
		V+D_{a}^{2}\ln V+(D_{a}\ln 
		V)^{2}\right)
	\label{F}
\end{equation}
as well as 
\begin{equation}
	C_{1}:=\frac{D_{a}'\Theta^{*}}{2D_{a}\Theta^{*}}
	\label{C1}
\end{equation}
which leads to (\ref{kalla2V}) in the form
\begin{equation}
	D_{a}\ln \Theta_{ba}=\frac{F}{D_{a}D_{b}\ln V}-C_{1}
	\label{kalla2a}.
\end{equation}
In addition we put 
\begin{equation}
		\begin{split}
		G:=&D_{b}\left(\frac{1}{6}D_{a}''\ln 
		V
		+\frac{1}{2}D_{a}D_{a}'\ln V+\frac{1}{2}D_{a}'\ln VD_{a}\ln V
		\right.\\
		&\left. +\frac{1}{6}D_{a}^{3}\ln V
		+\frac{1}{2}D_{a}^{2}\ln VD_{a}\ln V
		+\frac{1}{6}(D_{a}\ln V)^{3}
		\right)
	\end{split}
	\label{G}
\end{equation}
and
\begin{equation}
	C_{2}=\frac{D_{a}''\Theta^{*}}{6D_{a}\Theta^{*}}
		+\frac{D_{a}^{3}\Theta^{*}}{6D_{a}\Theta^{*}}.
	\label{C2}
\end{equation}
with which (\ref{pillet1aV}) takes the form
\begin{equation}
	G= D_{a}D_{b}\ln V\left(C_{2}+C_{1}D_{a}\ln \Theta_{ba}+\frac{1}{2}D_{a}'\ln 
		\Theta_{ba}+\frac{1}{2}D_{a}^{2}\ln 
		\Theta_{ba}+\frac{1}{2}(D_{a}\ln\Theta_{ba})^{2}\right) 
	\label{pillet1a2}.
\end{equation}

Eliminating $\Theta_{ba}$ with (\ref{kalla2a}) from (\ref{pillet1a2}) 
leads in a first step to
\begin{equation}
	\frac{G}{D_{a}D_{b}\ln V}-\frac{1}{2}D_{a}\left(\frac{F}{D_{a}D_{b}\ln V}\right)
	-\frac{F^{2}}{2(D_{a}D_{b}\ln V)^{2}}=C_{2}-\frac{C_{1}^{2}}{2}
	+\frac{1}{2}D_{a}'\ln\Theta_{ba}
	\label{pillet1a3}.
\end{equation}

Differentiating with respect to $D_{a}$ (note that the derivatives of 
the odd theta functions in $c_{1}$, $c_{2}$ vanish), we obtain with (\ref{kalla2a})
\begin{equation}
	D_{a}\left(\frac{G}{D_{a}D_{b}\ln V}-\frac{1}{2}D_{a}\left(\frac{F}{D_{a}D_{b}\ln V}\right)
	-\frac{F^{2}}{2(D_{a}D_{b}\ln V)^{2}}\right)=\frac{1}{2}
	D_{a}'\left(\frac{F}{D_{a}D_{b}\ln V}\right)
	\label{pillet2}.
\end{equation}
We have with $U:=D_{b}\ln V$
\begin{equation}
	\begin{split}
	&\frac{G}{D_{a}D_{b}\ln V}-\frac{1}{2}D_{a}\left(\frac{F}{D_{a}D_{b}\ln V}\right)
	-\frac{F^{2}}{2(D_{a}D_{b}\ln V)^{2}}\\
	&=\frac{D_{a}''U}{6D_{a}U}+\frac{D_{a}'D_{a}U}{4D_{a}U}
	+\frac{1}{2}D_{a}'\ln V\\
	&-\frac{D_{a}^{3}U}{12D_{a}U}+\frac{(D_{a}^{2}U)^{2}}{8(D_{a}U)^{2}}-\frac{(D_{a}'U)^{2}}{8(D_{a}U)^{2}},
	\end{split}
	\label{pillet2a}.
\end{equation}

Thus we get for (\ref{pillet2}) 
 	\begin{equation}
 	\begin{split}
 		0&=\frac{D_{a}''D_{a}U}{6D_{a}U}-\frac{D_{a}''UD_{a}^{2}U}{(6D_{a}U)^{2}}
 		-\frac{D_{a}^{4}U}{12D_{a}U}+\frac{D_{a}^{3}UD_{a}^{2}U}{3(D_{a}U)^{2}}\\
 		&-\frac{(D_{a}^{2}U)^{3}}{4(D_{a}U)^{3}}+\frac{(D_{a}'U)^{2}D_{a}^{2}U}{4(D_{a}U)^{3}}
 		-\frac{(D_{a}')^{2}DU}{4D_{a}U}.
 \end{split}
 	\label{pillet3U}
 \end{equation}
identical to equation (\ref{pillet3}) which 
concludes the proof.  Note that there are terms in (\ref{pillet2}) 
without a derivative $D_{a}$, but remarkably these terms all cancel 
leaving (\ref{pillet2a}) a relation for terms all involving this 
derivative. 

\section{Applications to integrable PDEs}
In this section we will  apply relation (\ref{pillet3}) to integrable 
equations, prove the second part of the main theorem and consider 
various reductions on special Riemann surfaces as known for the KP case. 

\subsection{Main theorem Part II}
To prove the second part of the main theorem, we 
define the function \( \phi(x,y,t) := \mathrm{D}_b \ln{\Theta^*_{ab} \Theta 
(x\mathbf{v}_{0}(a)+y \mathbf{v}_{1}(a)+ t 
\mathbf{v}_{2}(a)+\mathbf{d} }) \) 
and show that it solves the Schwarzian KP equation (\ref{SKP}). 

With our previous notations $ \phi = D_{b}\ln V=U$, moreover we can identify: $D_{a}=\partial_{x}$, $D_{a}'=\partial_{y}$ 
and $ D_{a}''=\partial_{t}$. Inserting the function \( \phi \) into (\ref{SKP}) we get:
\begin{align*}
    \frac{D_{a}''D_{a} U}{D_{a} U}-\frac{D_{a}^2 U}{(D_{a} 
	U)^2}D_{a}'' U-\frac{1}{2}\frac{D_{a}^4 U}{D_{a} 
	U}+\frac{2}{(D_{a} U)^2}D_{a}^2 U D_{a}^3 U \\
	-\frac{3}{2(D_{a} U)^{3}}(D_{a} U)^{3}+\frac{3}{2}\frac{D_{a}^2 
	U}{(D_{a} U)^{3}}(D'_{a} U)^{2}-\frac{3}{2}\frac{(D'_{a})^2 
	U}{D_{a} U}=0
\end{align*}
which is equivalent to (\ref{pillet3}) thus proving this part of the 
theorem. 
\begin{remark}
	It is well known, see for instance \cite{Mum,BBEIM}, that one way 
	to represent meromorphic functions on a Riemann surface is in 
	terms of second logarithmic derivatives of theta functions. The 
	function $D_{b}\ln \Theta^{*}_{ab}\Theta$ is a priori not 
	independent of the path between $a$ and $b$ in 
	$\int_{a}^{b}\mathrm{d}\omega$. Both points have to be in 
	the same fundamental polygon. A possibility to avoid this 
	condition would be to consider $U=\partial_{x}^{-1}u$, where $u:=D_{a}D_{b}\ln 
	V$ is path independent; the 
	anti-derivative is defined as 
	$\partial_{x}^{-1}=\int_{a}^{x}dx'$. 
\end{remark}

The Schwarzian KP equation 
can also be written in the form 
\begin{equation}
	\left(\frac{U_{t}}{U_{x}}-\frac{1}{2}\frac{U_{xxx}}{U_{x}}+\frac{3}{4}
	\frac{U_{xx}^{2}}{U_{x}^{2}}\right)_{x}+\frac{3}{2}\frac{U_{xx}}{U_{x}^{3}}U_{y}^{2}
	-\frac{3}{2}\frac{U_{yy}}{U_{x}}=0.
	\label{KP2}
\end{equation}
Integrating with respect to $x$, we get
\begin{equation}
	U_{t}-\frac{1}{2}U_{xxx}+\frac{3}{4}\frac{U_{xx}^{2}-U_{y}^{2}}{U_{x}} +\frac{3}{2}U_{x} \partial_{x}^{-1}\left(\frac{U_{xy}U_{y}}{U_{x}^{2}}
	-\frac{U_{yy}}{U_{x}}\right)
	\label{KP3},
\end{equation}
which can be written in the form 
\begin{equation}
	U_{t}-\frac{1}{2}U_{xxx}+\frac{3}{4}\frac{U_{xx}^{2}-U_{y}^{2}}{U_{x}} -\frac{3}{2}U_{x} W_{y}
	\label{KP4},
\end{equation}
where 
$W_{x}:=U_{y}/U_{x}$. This is equation (13) in \cite{BK} after the 
change of time $t\mapsto 2t$. 

\subsection{Reductions on special Riemann surfaces}

Let us restrict our attention to the special case of the  Riemann 
surface \( \mathcal{R} \) being  hyperelliptic, i.e., given by the 
zero locus of the polynomial \( P(\lambda, \mu) = \mu^2-\prod_{j=1}^N 
(\lambda-\lambda_j) \) where \( \lambda_j \in \mathbb{C} \), 
$j=1,\ldots,N$, and \( N=2g+1 \) or \( N=2g+2 \), and denote by \( \pi: \mathcal{R} \longrightarrow \mathbb{CP}^1 \) the projection onto the Riemann sphere. \\
If \( a \) is a branch point of \( \pi \) then the hyperelliptic 
involution locally reads \( \sigma: k_a \mapsto -k_a \) and its 
pullback on \( \omega_j \) is given by: \( \sigma^* \omega_j = 
-\omega_j \), hence the Taylor expansion of \( \omega_j \) around \(a 
\) must be even in \(k_a \), i.e \( \mathbf{v}_{1} =0 \) and thus \( \mathrm{D_a'}=0 \). 
Eliminating $\Theta_{ba}$ from (\ref{pillet1a2}) via 
(\ref{F}), we get 
\begin{cor}
	 Identity (\ref{pillet3}) reduces on hyperelliptic surfaces with 
$a$ being a branch point to
\begin{equation}
	0=\frac{1}{6}D_{a}''D_{b}\ln V - \frac{1}{12}D_{a}^{3}D_{b}\ln 
	V-  C_{2}D_{a}D_{b}\ln V +\frac{1}{8}\frac{(D_{a}^{2}D_{b}\ln 
	V)^{2}}{D_{a}D_{b}\ln V}
	\label{red1}.
\end{equation}
	
\end{cor}

If we put again $U=D_{b}\ln V$, $D_{a}''=\partial_{t}$ and 
$D_{a}=\partial_{x}$, we get for (\ref{red1})
\begin{equation}
	\frac{1}{6}U_{t}-\frac{1}{12}U_{xxx}-C_{2}U_{x}+\frac{1}{8} 
	\frac{U_{xx}^{2}}{U_{x}}=0
	\label{red2}.
\end{equation}
This implies
\begin{cor}
The function \( \phi(x,t) := \mathrm{D}_b \ln{\Theta^*_{ab} \Theta 
(x\mathbf{v}_{0}(a)+ t \mathbf{v}_{2}(a)+\mathbf{d} }) \) solves the Schwarzian KdV equation:
\begin{equation}
    \frac{1}{6}\frac{\phi_t}{\phi_x}-\frac{1}{12} \{ \phi ; x \} = 
	C_{2}
    \label{SKdV}.
\end{equation}
\end{cor}

As in Chapter 3.4 of \cite{BBEIM} for KP, there is also a reduction 
to a Schwarzian Boussinesq equation. If the surface $\mathcal{R}$ is 
given by a trigonal curve, i.e., a curve on which a meromorphic 
function with a third order pole at a point $a\in \mathcal{R}$ and no 
other singularities exists. A simple example of such a curve is
\begin{equation}
	\mu^{4}=\prod_{i=1}^{4}(\lambda-E_{i}).
	\label{trig}
\end{equation}
In this case $D_{a}''=0$ which leads for (\ref{pillet3}) to
\begin{equation}
		0=
		-\frac{D_{a}^{4}U}{12D_{a}U}+\frac{D_{a}^{3}UD_{a}^{2}U}{3(D_{a}U)^{2}}\\
		-\frac{(D_{a}^{2}U)^{3}}{4(D_{a}U)^{3}}+\frac{(D_{a}'U)^{2}D_{a}^{2}U}{4(D_{a}U)^{3}}
		-\frac{(D_{a}')^{2}DU}{4D_{a}U}.
	\label{pillet3b}
\end{equation}

The function
\( \phi(x,t) := \mathrm{D}_b \ln{\Theta^*_{ab} \Theta 
(x\mathbf{v}_{0}(a)+t \mathbf{v}_{1}(a)+\mathbf{d} }) \) then gives a 
solution to the Schwarzian Boussinesq equation \cite{WeissI}
\begin{equation}
    \Big( -\frac{1}{2} \{ \phi ; x \} \Big)_x - \frac{3}{2} \Big( 
	\frac{\phi_t}{\phi_x} \Big)_t-\frac{3}{4} \Big( \frac{\phi_t^2}{\phi_x^2} \Big)_x=0
    \label{SB}
\end{equation}

\section{Conclusion}
In this paper we have studied degenerations of Fay's identities in higher order of the 
local parameter $\tau$ near one of the points. The starting point was Fay's 
identity for 3 points on a Riemann surface in the form (\ref{Fay1V}). 
The case of order $\tau^{3}$ was studied in detail as well as its 
application to the Schwarzian KP equation. It appears straight 
forward to generalize this approach to higher orders of the parameter 
$\tau$. A standard Taylor expansion of the quantity $V(c)$ yields for 
the left hand side of equation (\ref{Fay1V}) 
\begin{equation}
	D_{b}\sum_{m=1}^{\infty}\frac{1}{m!V}\left(\tau 
	D_{a}+\frac{\tau^{2}}{2}D_{a}'+\frac{\tau^{3}}{6}D_{a}''+\ldots 
	\frac{\tau^{k}}{k!}D_{a}^{(k)}+\ldots\right)^{m}V
	\label{Vex}.
\end{equation}
Thus  one gets in order $\tau^{n}$
\begin{equation}
	D_{b}\left(\frac{D_{a}^{n}V}{n! 
	V}+\frac{nD_{a}^{n-1}D_{a}'V}{2(n-1)!V}+\ldots+\frac{D_{a}^{(n)}V}{n!V}\right)
	\label{Vn}.
\end{equation}
The same expansion can be obtained on the right hand side of 
(\ref{Fay1V}) for $\Theta^{*}_{ac}$, where all even order derivatives 
vanish for symmetry reasons, and for $\Theta_{bc}$ leading to 
derivatives of $\Theta_{ba}$. As in section 3, the latter terms  can be 
replaced via (\ref{kalla2a}) by derivatives of $\Theta$. As in 
(\ref{pillet2}), it will be necessary to differentiate with respect 
to $D_{a}$ in general in order to eliminate all terms with 
$\Theta_{ba}$. It is beyond the scope of the current paper to detail 
the resulting relations and to establish a potential relation to 
integrable PDEs and whether these are from a hierarchy of Schwarzian 
KP equations. An interesting question is also whether a similar 
approach can be applied to the degeneration of identity (\ref{Fay2}) 
in the limit $b\to a$
in higher orders of the local parameter near $a$, which would lead to 
a generalisation of relation (\ref{Fay3}). This will be the subject 
of future research. 

Another interesting aspect would be to relate the present work to the 
bilinear approach studied in \cite{BK2}. In this article the authors 
describe the tau and Baker-Akhiezer functions associated to some generalized KP hierarchy (the Schwarzian KP hierarchy being one of them). Their derivation is based on a generalized Hirota's bilinear identity based on the equation:
\begin{equation*}
    \int_{\partial G} \chi( \nu, \mu ; g_1) g_1(\nu) g_2^{-1}(\nu) \chi(\lambda, \nu ; g_2) d \nu =0
\end{equation*}
Where (following the notations of \cite{BK2})  \( G \) is the unit disk, \( (\lambda, \mu) \in \mathbb{C}^2 \) are spectral parameters, \( g_1(\nu) = g(\nu, \mathbf{x})=\exp{\big( \sum_{i=1}^{\infty} x_i \nu ^{-i} \big) } \), \( g_2(\nu)= g(\nu, \mathbf{x'}) \) and \( \chi(\lambda, \mu) \) is an unknown meromorphic function in both variables. \\
The associated tau function is given by:
\begin{equation*}
    \chi(\lambda, \mu, \mathbf{x}) = \frac{1}{(\lambda-\mu)}\frac{\tau(\mathbf{x}-[\lambda]+[\mu])}{\tau(\mathbf{x})}
\end{equation*}
where \( \mathbf{x}+[\mu]=x_i+[\mu]_i \), with \( [\mu]_i=\frac{1}{i} \mu^i \), \( 0 \leq i < \infty \).
They proved that this tau function satisfies the following addition formula for \(a\), \(b\), \(c\) and \(d\) some arbitrary complex numbers :
\begin{equation*}
\begin{split}
    & (a-c)(d-b) \tau(\mathbf{x}+[a]+[c])\tau(\mathbf{x}+[d]+[b])+ \\
    & +(d-a)(b-c)\tau(\mathbf{x}+[d]+[a])\tau(\mathbf{x}+[b]+[c])+\\
    & +(b-a)(d-c)\tau(\mathbf{x}+[b]+[a])\tau(\mathbf{x}+[d]+[c])=0
\end{split}
\end{equation*}
As established in the seminal paper \cite{Shio}, this addition formula is nothing more than Fay's trisecant identity when a tau function can be written in terms of Riemann theta functions. All these considerations lead to the following conjecture:
\begin{conj}
There exists a quadratic form \( \mathbf{Q}(\mathbf{t}):= \sum_{i,j=1}^4 Q_{ij}t_i t_j \), \( Q_{ij} \in \mathbb{C} \), such that the function
\begin{equation*}
    \tau(\mathbf{t}):=\exp{ (\mathbf{Q}(\mathbf{t})) }\Theta(t_1\mathbf{v}_{0}(a)+t_2 \mathbf{v}_{1}(a)+ t_3 
\mathbf{v}_{2}(a) + t_4 \mathbf{v_0}(b)+ \mathbf{d})
\end{equation*}
is a tau function for the Schwarzian KP equation \ref{SKP} in the sense of \cite{BK2}, the variables \( t_1 \), \( t_2 \), \( t_3 \) being respectively identified with \( x \), \(y \), \(t \) and \(t_4 \) is an auxiliary parameter.
\end{conj}
This conjecture relies on the fact that the Schwarzian KP hierarchy 
is generated in a very similar fashion as the classical KP hierarchy \cite{BK2}. Hence a tau function for the full Schwarzian KP hierarchy could be of the form \( \Tilde{\tau} := \exp{ (\mathbf{Q}(\mathbf{t})) } \Theta( \mathbb{V} \mathbf{t} + \mathbf{d}) \), with \( \mathbb{V} \) a \( g \times \infty \) matrix whose columns are the vectors associated to the different "time" variables, and the higher order degenerations discussed above could have an explicit and compact form when expressed with Hirota's symbols.

\appendix
\section{Interesting links with other integrable systems}
To allow for an independent reading, we collect in this appendix some 
facts on the Schwarzian PDEs appearing naturally in the context of higher order 
degenerations of Fay's identity. 
The Schwarzian KP, KdV and Boussinesq equations originally appear in 
a series of papers by Weiss \cite{WeissI}, \cite{WeissII}, in an 
extensive study of the Painlev\'e property of many integrable PDEs. 
These three equations were shown to be in some sense prototypical 
PDEs satisfying this property. They are also linked to some 
integrable systems appearing in physics through B\"acklund and Miura 
transforms. We would like to explain these relationships in this section. 

First we can observe that these equations are invariant under a 
M\"obius transformation, in the sense that if the function \( \phi \) 
solves the Schwarzian KP, Boussinesq or KdV equations then \( \psi := 
\frac{A \phi +B}{C \phi +D} \) with \( AD-BC \neq 0 \) is also a solution of these equations. Moreover this transformation plays the role of the B\"acklund transform for these three equations. \\
If \( \phi(x,y,t) \) solves the Schwarzian KP equation \( \Big( \frac{\phi_t}{\phi_x}+ \{ \phi ; x \} \Big)_x + \Big( \frac{\phi_y}{\phi_x} \Big)_y + \frac{1}{2} \Big( \frac{\phi_y^2}{\phi_x^2} \Big)_x=0 \) then the Bäcklund transformed function \( u(x,y,t) := 12 \frac{\partial^2}{\partial x ^2} \ln{\phi} + v  \) with \( v= 3\frac{\phi_{xx}^2}{\phi_x^2} -4 \frac{\phi_{xxx}}{\phi_x} - \frac{\phi_t}{\phi_x}-\frac{\phi_{y}^2}{\phi_x^2} \) is a solution of the KP equation \( u_{tx}+u_x^2+uu_{xx}+u_{xxxx}+u_{yy}=0 \).  

Similarly the Schwarzian KdV equation is related to the classical KdV 
equation via the same type of B\"acklund transformation: let \( 
\phi(x,t) \) be a solution of  \( \frac{\phi_t}{\phi_x}- \{ \phi ; x 
\} = \lambda\) then \( u= 12 \frac{\partial^2}{\partial x ^2} 
\ln{\phi} + v \) is a solution of \( u_t + u_{xxx}+uu_x = 0 \) 
where \( v \) is given by \( v= 3 \frac{\phi_{xx}^2}{\phi_x^2} -4 
\frac{\phi_{xxx}}{\phi_x} - \frac{\phi_t}{\phi_x} \) (the very same Bäcklund transformation relates solutions of the Schwarzian Boussinesq equation to solutions of the classical Boussinesq equation). Moreover if we set \( \lambda = 0 \) then \( w := 2 \frac{\phi_x}{\phi}+\frac{\phi_{xx}}{\phi_x} \) solves the modified KdV equation \( w_t+ \frac{\partial}{\partial x} \big( w_{xx}-\frac{w^3}{2} \big) =0 \).\\
Finally, under the change of variables (Miura transform):
\begin{equation*}
    x \to \phi, \quad t \to t, \quad \phi \to x
\end{equation*}
which implies
\begin{equation*}
    \{ \phi ; x \}= - \phi_x^2 \{ x ; \phi \}, \quad 
	\phi_x=\frac{1}{x_{\phi}}, \quad x_t = - \frac{\phi_t}{\phi_x}.
\end{equation*}
Therefore, under this change of variables, the Schwarzian KdV equation becomes
\begin{equation*}
    x_{\phi}^2 x_t = \lambda x_{\phi}^2 + \{ x ; \phi \},
\end{equation*}
or replacing the Schwarzian derivative \( \{ x ; \phi \} \) by its 
explicit expression,
\begin{equation*}
    x_t = \lambda - \frac{1}{2} \Big( \frac{1}{x_{\phi}} \Big)_{\phi 
	\phi} + \frac{3}{2} \Big( \frac{1}{x_{\phi}} \Big)_{\phi}^2.
\end{equation*}
With \( v := x_{\phi}^{-1} \), the previous equation is equivalent to
\begin{equation*}
    v_t=v^{3} v_{\phi \phi \phi}
\end{equation*}
i.e. the Dym equation.

\end{document}